# Understanding the drivers of sustainable land expansion using a patch-generating land use simulation (PLUS) model: A case study in Wuhan, China

Xun Liang[a,b,c], Qingfeng Guan[a,b*], Keith C. Clarke[c], Shishi Liu[d], Bingyu Wang[e], Yao Yao[a,b],

[a] School of Geography and Information Engineering, China University of Geosciences, Wuhan, Hubei 430078, China

[b] National Engineering Research Center of GIS, China University of Geosciences, Wuhan 430078, Hubei province, China

[c] Department of Geography, University of California, Santa Barbara, Santa Barbara, CA 93106-4060, United States of America

[d] College of Resources and Environment, Huazhong Agricultural University, Wuhan, Hubei 430070, China

[e] Department of Natural Environmental Studies, Graduate School of Frontier Sciences, The University of Tokyo, Japan.

* Corresponding author. E-mail address: guanqf@cug.edu.cn (Qingfeng Guan).

**Abstract**

Cellular Automata (CA) are widely used to model the dynamics within complex land use and land cover (LULC) systems. Past CA model research has focused on improving the technical modeling procedures, and only a few studies have sought to improve our understanding of the nonlinear relationships that underlie LULC change. Many CA models lack the ability to simulate the detailed patch evolution of multiple land use types. This study introduces a patch-generating land use simulation (PLUS) model that integrates a land expansion analysis strategy and a CA model based on multi-type random patch seeds. These were used to understand the drivers of land expansion and to investigate the landscape dynamics in Wuhan, China. The proposed model achieved a higher simulation accuracy and more similar landscape pattern metrics to the true landscape than other CA models tested. The land expansion analysis strategy also uncovered some underlying transition rules, such as that grassland is most likely to be found where it is not strongly impacted by human activities, and that deciduous forest areas tend to grow adjacent to arterial roads. We also projected the structure of land use under different optimizing scenarios for 2035 by combining the proposed model with multi-objective programming. The results indicate that the proposed model can help policymakers to manage future land use dynamics and so to realize more sustainable land use patterns for future development. Software for PLUS has been made available at https://github.com/HPSCIL/Patch-generating_Land_Use_Simulation_Model

# 1. Introduction

Since Tobler first applied cellular automata (CA) to geographic modeling (Tobler,1979), CA have been extensively applied to model spatiotemporal land use dynamics under the influences of natural and socioeconomic factors, including their interactions at different scales (Chen et al., 2020; Clarke & Gaydos, 1998). Ideally, CA models are developed and applied to inform land use policy and land management decisions and their results should be clear and accessible during the decision-making process (Guzman, Escobar, Peña, & Cardona,2020; Sohl & Claggett,2013). However, most CA modeling research has focused on the improvement of technical modeling procedures or on model calibration and rules, and little attention has been paid to the need for a more conceptual understanding of the underlying causes of LULC (Cao, Tang, Shen, & Wang, 2015; Engelen, White, Maarten, & Bernhard, 2002). Sohl & Claggett (2013) noted that CA models have not been widely used for planning and policy development because CA modelers tend to focus on exercises based on scientific inquiry, while decision-makers focus on outcomes and strategy-driven exercises (Leenhardt et al., 2012). Recently some studies have improved the application of CA models in planning and decision making by using them to create urban growth boundaries (Huang, Huang, & Liu, 2019; Liang et al.,2018), to design urban forms for development zones (Liang et al., 2020) or to explore the effects of greenbelt elimination (Park, Clarke, Choi, & Kim, 2017). However, for decision-makers, many of the existing CA models are still: (1) weak at revealing the underlying drivers of land use change (Sohl & Claggett, 2013); and (2) unable to spatio-temporally capture the evolution of patches of multiple land uses, especially for the patch evolution of natural land use types (Meentemeyer et al., 2013;Yang, Gong, Tang, & Liu, 2020). The reasons and mechanisms that result in these deficiencies are described below.

## 1.1. Discovering the underlying drivers of LULC change

The rule mining methods used for CA models are one reason for the lack of discovery of the underlying scientific basis of LULC. The most commonly used strategy for CA models is to capture the relationships between revealed land use transitions and multiple explanatory driving factors into the CA behavior rules (Pijanowski, Pithadia, Shellito, & Alexandridis, 2005; Shu et al.,2017), what might be called a transition analysis strategy (TAS). The TAS must extract sample cells with land use states that are changed during the time period between two dates of land use data to examine the neighboring states, or distances and weights of the contributing factor layers (Figure 1(a)). This kind of CA model includes logistic-CA (Chen, Li, Liu, & Ai, 2013; Wu & Webster,1998), ANN-CA (Omrani, Parmentier, Helbich, & Pijanowski, 2019), and optimization-based CA (Feng, Liu, Tong, Liu, & Deng, 2011). However, when simulating the transitions of multiple land use types, CAs based on a TAS become very difficult to implement. For example, if a region has $K$ land use types, there are a total of $K^2 - K$ type transitions (Figure 1(a)), which means that the number of transition types theoretically increases exponentially with the number of land use types. Analyzing all transition types increases the computational complexity of the model structure of CA and reduces the model's flexibility and general applicability.

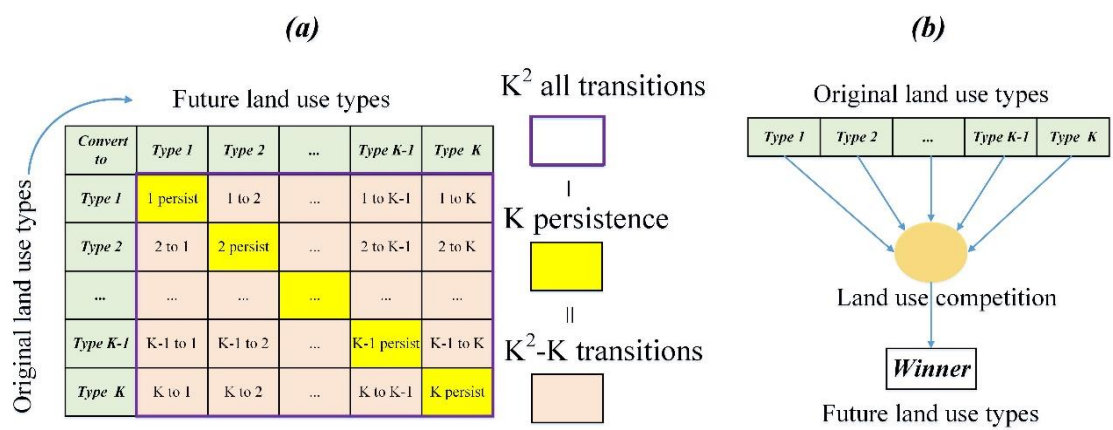

Figure 1. Panel (a) is the matrix of the transition types generated by the transition analysis strategy; panel (b) shows the competition in the simulation model based on a pattern analysis strategy.

To solve this problem, researchers have developed another rule mining method based on an analysis of the patterns of land use instead of the changes in the type of land use (Verburg, Schot, Dijst, & Veldkamp, 2004), naming this method the pattern analysis strategy (PAS). CA models based on PAS calculate the probability of occurrences of land use types in each cell with just one date for the land use data. They determine the future cell states through competition between the land use types, which avoids analyzing transition type combinations that can increase exponentially with the number of land use types in the process of mining transition rules (Figure 1(b)). Many simulation models are designed based on PAS, such as the CLUE-s model (Verburg et al., 2002), Fore-SCE model (Sohl & Sayler, 2008) and the FLUS model (Liu et al., 2017). However, CA models based on PAS inherently lack the ability to reveal how the driving factors cause land use change (Amato, Pontrandolfi, & Murgante, 2015) because they are unable to analyze the evolution of land use from the perspective of class transitions. Further, the distribution rules of PAS are not as valuable as transition rules, since the distribution rules based on one land use map cannot capture the rules of land use change over a specific time interval. In summary, although previous CA models based on TAS and PAS can provide valuable model outcomes, such as forecasting land use patterns under climate scenarios (Chen et al., 2020), estimating the impacts of land use change on environmental variables (Ahmed, Kamruzzaman, Zhu, Rahman, & Choi, 2013), or testing the effects of planning policies on urban growth (Liang, Liu, Li, Zhao, & Chen, 2018), nevertheless these models are unable to provide insights concerning the drivers behind the dynamics of individual land types transitions and the strengths of their effects on the expected changes.

### 1.2. An obstacle to simulating landscape evolution

A more complete understanding of local land use dynamics is of primary importance for effective land use planning and decision making. Many models that have relied on cell state transitions have struggled to generate realistic spatial structures at the detailed scales that are appropriate for plan-specific decision-making (Meentemeyer et al., 2013). Therefore, the development of patch growth simulation models has received increasing attention (Meentemeyer et al. ,2013;Yang et al., 2020). Simulating the succession of land use patches can be used to detect ecologically significant changes and to assess, design, and plan ecosystem management activities in advance (Keane, Parsons, & Hessburg, 2002). Several studies have used patch-based landscape metrics to examine past land use trends, and to compare them with simulated future behavior (Herold, Goldstein, & Clarke, 2003).

In the last few years, researchers have developed a set of CA models or mechanisms (and even vector-based CAs) to simulate urban growth based on patch units (Yao et al., 2017). However, vector-based CA cannot be used to simulate the dynamics of natural land use types (e.g., forest and grassland) because the parcels of natural land use types can be very small, irregular and dispersed and not as well-organized as urban land use types. Hence almost all the vector-based CAs are applied to urban land use types (e.g., residential land and industrial land). Vector-based land use data is more difficult to obtain than raster-based land use data, which is an important reason to constrain the application of vector-based CA, especially in large scale regions. Some previous studies have highlighted the importance of simulating the dynamics of the patches using a raster-based CA model which can simulate the growth of urban patches, for example, the SLEUTH model (Herold et al., 2003) and the Patch-Logistic-CA (Chen et al., 2013). However, to the best way to simulate the patch spatio-temporal dynamics of multiple land use types is still unclear. Sohl, Sayler, Drummond, & Loveland (2007) promoted a patch growth land use simulation by proposing the raster-based Fore-SCE model. However, the Fore-SCE model separately processes the urban land with a simple density slicing technique because it cannot address the cases where there are very large numbers of very small patches, which loses the temporal dynamics and the ability to simulate the synchronous evolution of multiple and multi-type patches. In short, previous studies lack flexible patch-based strategies for modeling the patch growth of multiple natural land use types at a fine-scale resolution, and the use of these CA models is limited for actual planning, decision-making, or policymaking purposes.

To promote better understanding of the relationships underlying LULC change and to improve patch-growth simulation, in this study, we developed a raster-based patch-generating land use simulation (PLUS) model to help reveal the underlying drivers and their differing contributions to change. The model includes a new data mining framework for identifying the rules of land use change, which combines the advantages and overcome the disadvantages of both TAS and PAS. We propose an approach that combines a CA model with a patch-generating simulation strategy to improve the model's ability to emulate and simulate the real landscape pattern. Multiple objective programs were used to generate a case study of a sustainable land use scenario and other scenarios using the PLUS model. The proposed model was used to explore the underlying causes of land use change in Wuhan, China, from 2000–2013. Next, the model was used to identify sustainable land use scenarios for Wuhan from 2013 to 2035. A software package for PLUS has been made available at https://github.com/HPSCIL/Patch-generating_Land_Use_Simulation_Model which provides a user-friendly interface for potential users.

## 2. Method

The PLUS model contains two modules: (1) a rule-mining framework based on a land expansion analysis strategy (LEAS); and (2) a CA based on multi-type random patch seeds (CARS). Multiple-objective programming (MOP) was used to determine the optimal land use structures under different scenarios. The general structure of the simulation framework of this study is illustrated in Figure 2.

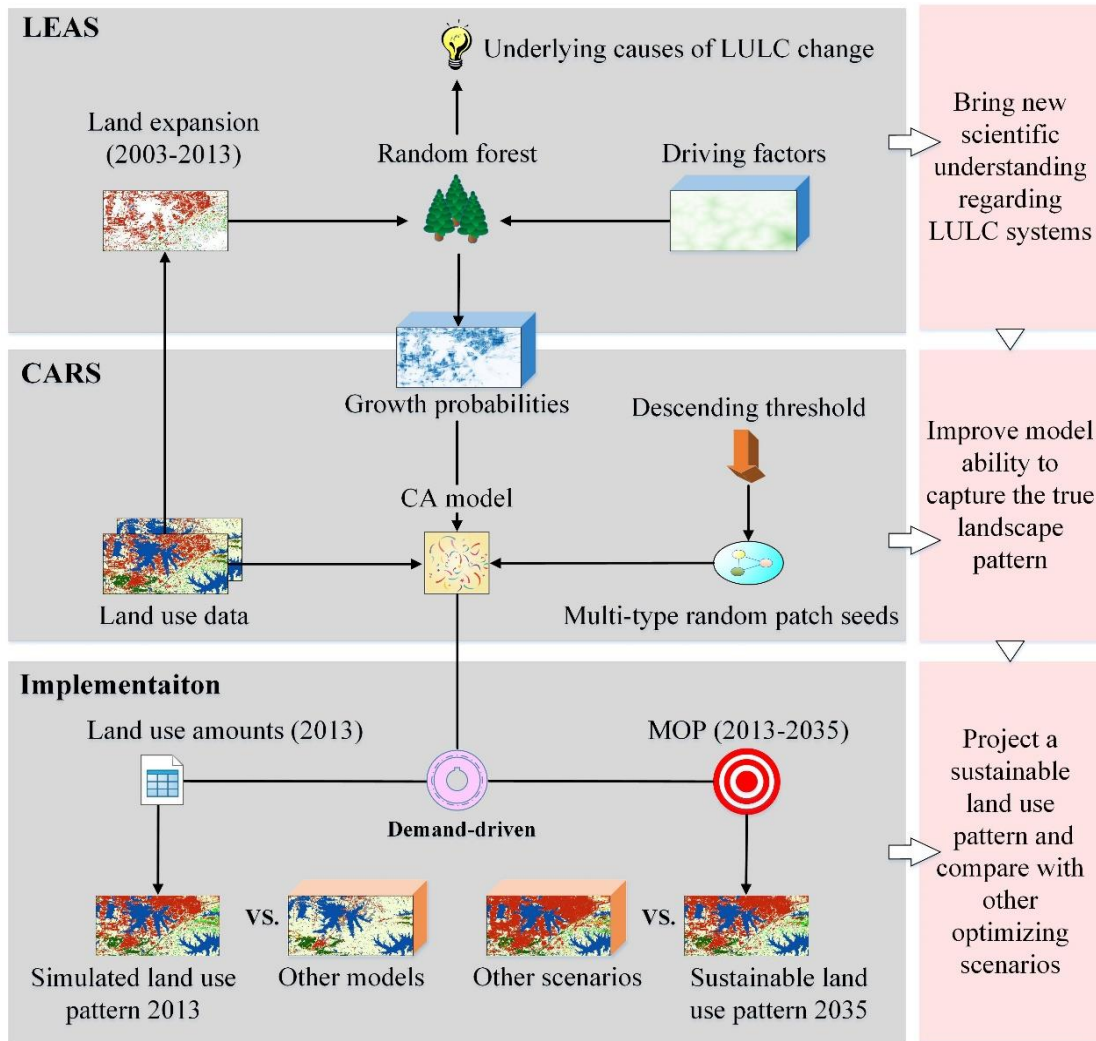

Figure 2. The framework of the proposed PLUS model and the modeling workflow.

## 2.1. Rule-Mining Framework based on Land Expansion Analysis Strategy

2.1.1. Land Expansion Analysis Strategy

The LEAS proposed in this study requires two dates of land use data as its TAS. We overlaid the two periods of land use data and extracted the cells with changed states from the later date of the land use data, which represented the change regions for each land use type. Sampling points were randomly selected, and were divided into subsets according to their land use types, which were analyzed separately using a data mining method (Figure 3). For example, when we mined the relationship between the expansion of a land use type and the driving factors, the labels of the samples of the expansion of this type were set to '1', and the labels of other samples were set to '0'. Thus, the training dataset can be reconstructed using the marked labels and the values of the multiple driving factors at the same locations extracted. Next, the training dataset was used to train the data mining algorithm to obtain the transition rules for each land use type.

By using LEAS, we can obtain the transition rules for all land use types by analyzing the growing patches of each changed land use ("to" land use type) while ignoring their source types ("from" land use type), which avoids the analysis of transition types that increase exponentially in number with the number of land use types and effectively simplifies the analysis procedure for land use change. The transition rules obtained from the LEAS have a temporal aspect, which has the ability to describe the properties of land use change during a specific time interval. Moreover, this framework also provided new perspectives for analyzing the drivers of land use change, which are described in section 4.2.

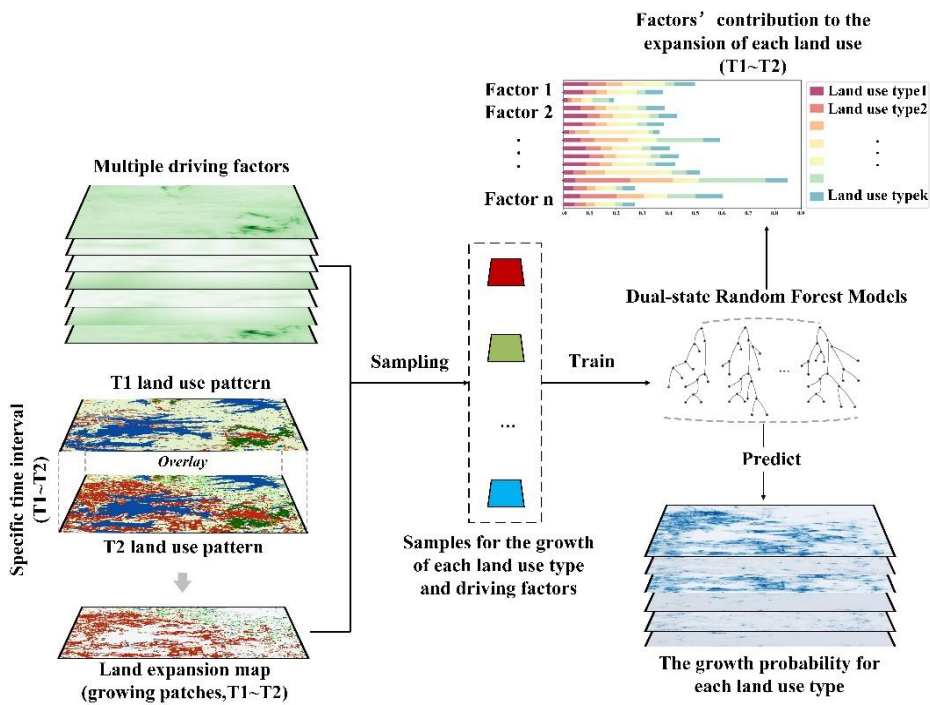

Figure 3. Workflow of the rule mining framework based on a land expansion analysis strategy with random forest models.

### 2.1.2. Random Forest Classification for Dual-state decisions

The LEAS transforms the mining of transition rules of each land use type into a binary classification problem (obtain the change and inertia probabilities of each land use type), which can be solved with many data mining methods. This study employed a random forest classification (RFC) algorithm to explore the relationships between the growth in each land use type and the multiple driving factors. The RFC algorithm is a decision tree-based ensemble classifier built from each sub-dataset that extracts random samples from the original training dataset. RFC algorithms are able to process high-dimension data as well as dealing with multicollinearity among variables, and finally output the growth probability $P_{i,k}^d$ of land use type $k$ at cell $i$.

$$P_{i,k}^d(x) = \frac{\sum_{n=1}^{M} I(h_n(x)=d)}{M} \tag{Eq. 1}$$

The value of $d$ is either 0 or 1; a value of 1 indicates that there were other land use types that changed to land use type $k$, while 0 represents other transitions; $x$ is a vector that consists of multiple driving factors; $I(\cdot)$ is the indicative function of the decision tree set; $h_n(x)$ is the prediction type of the $n-$th decision tree for vector $x$; and $M$ is the total count of decision trees. In addition, the RF algorithm has the advantage of measuring the importance of independent variables to the variation of dependent variables, which can be calculated according to the variation of the out-of-bag error caused by stochastic noise. This method has been implemented in many studies (Yao et al., 2017)

### 2.2. CA Model based on Multi-type Random Patch Seeds

The CARS module is a CA model that includes a patch-generation mechanism based on multi-type random seeds of land uses (Figure 4). The CA model is a scenario-driven land use simulation model that integrates 'top-down' (i.e., global land use demands) and 'bottom-up' (i.e., local land use competition) effects. In the simulation process, the land use demands affect the local land use competition through a self-adaptive coefficient, driving the amount of land use to reach future demands.

2.2.1. Feedbacks between Macro Demands and Local Competition

The basic formula for calculating the overall probability ($OP_{i,k}^{d=1,t}$) of the land use type $k$ can be expressed as follows:

$$OP_{i,k}^{d=1,t} = P_{i,k}^{d=1} \times \Omega_{i,k}^t \times D_k^t \qquad \text{(Eq. 2)}$$

Where $P_{i,k}^{d=1}$ represents the growth probability of land use type $k$ at cell $i$; $D_k^t$ is the impact of the future demand for land use type $k$, which is a self-adaptive driving coefficient that depends on the gap between the current amount of land at iteration $t$ and the target demand of land use $k$; and $\Omega_{i,k}^t$ denotes the neighborhood effects of cell $i$, which are the cover proportions of the land use components of $k$ within the following neighborhood.

$$\Omega_{i,k}^t = \frac{con(c_i^{t-1}=k)}{n \times n - 1} \times w_k \qquad \text{(Eq. 3)}$$

where $con(c_i^{t-1} = k)$ represents the total number of grid cells occupied by land use type $k$ at the last iteration within the $n \times n$ window and $w_k$ is the weight among the different land use types because there are different neighborhood effects for the different land use types. The default value of $w_k$ is 1, but it can be defined by the model users. The self-adaptive method of $D_k^t$ is as follows:

$$D_k^t = \begin{cases} D_k^{t-1} & if\ |G_k^{t-1}| \leq |G_k^{t-2}| \\ D_k^{t-1} \times \frac{G_k^{t-2}}{G_k^{t-1}} & if\ 0 > G_k^{t-2} > G_k^{t-1} \\ D_k^{t-1} \times \frac{G_k^{t-1}}{G_k^{t-2}} & if\ G_k^{t-1} > G_k^{t-2} > 0 \end{cases} \qquad \text{(Eq. 4)}$$

where $G_k^{t-1}$ and $G_k^{t-2}$ are the differences between the current amount of, and future demand for, land use type $k$ at the $t-1th$ and $t-2th$ iteration. Finally, a roulette wheel is constructed according to the overall probabilities of all the land use types and used to select the land use state in the next iteration (Liu et al., 2017).

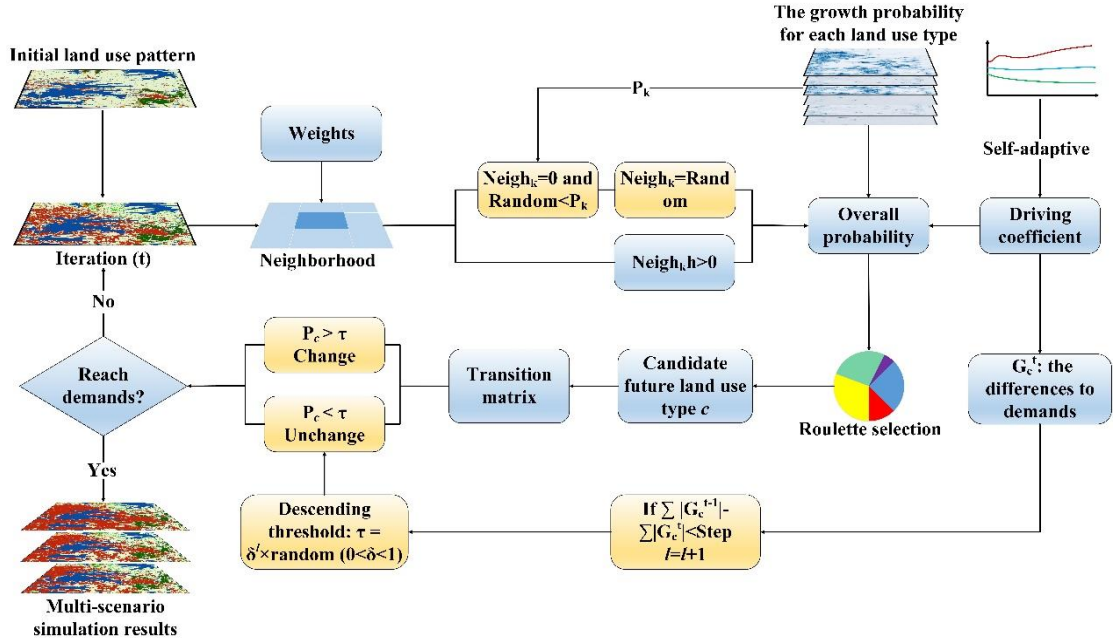

Figure 4. A schematic diagram of the CA model based on multi-type random patch seeds. The yellow boxes were the key steps in modeling the emergence and growth of multi-type patches.

2.2.2. Multi-type Random Patch Seeds based on a Descending Threshold

To simulate the patch evolution of multiple land use types, a multi-type random patch seeding mechanism based on threshold descent was used in this study, implemented through the calculation process of overall probability (Figure 4). This mechanism generated change 'seeds' on the growth probability surface ($P_{i,k}^{d=1}$) for each land use type through the use of a Monte Carlo approach when the neighborhood effects of a land use $k$ were equal to 0:

$$OP_{i,k}^{d=1,t} = \begin{cases} P_{i,k}^{d=1} \times (r \times \mu_k) \times D_k^t & if\ \Omega_{i,k}^t = 0\ and\ r < P_{i,k}^{d=1} \\ P_{i,k}^{d=1} \times \Omega_{i,k}^t \times D_k^t & all\ others \end{cases} \quad \text{(Eq. 5)}$$

where $r$ is a random value ranging from 0-1; $\mu_k$ is the threshold to generate the new land use patches for land use type $k$, which is determined by the model users. The seeds may generate a new land use type and grow into new patches formed by a set of cells with the same land use type. To control the generation of multiple land use patches, a threshold descending rule for the competition process was proposed to restrict both the organic and spontaneous growth of all land use types. If a new land use type wins in a round of competition, a decreasing threshold $\tau$ is employed to assess the candidate land use type $c$ that was selected by the roulette wheel as follows:

$$\text{If } \sum_{k=1}^{N}|G_c^{t-1}| - \sum_{k=1}^{N}|G_c^t| < Step \quad \text{Then, } l = l+1 \quad \text{(Eq. 6)}$$

$$\begin{cases} Change & P_{i,c}^{d=1} > \tau \text{ and } TM_{k,c} = 1 \\ No\ change & P_{i,c}^{d=1} \leq \tau \text{ or } TM_{k,c} = 0 \end{cases} \quad \tau = \delta^l \times r \quad \text{(Eq. 7)}$$

where $Step$ is the step size of the PLUS model to approximate the land use demand; $\delta$ is the decay factor of decreasing threshold $\tau$, which ranges from 0 to 1 and is set according to the expert; $r$ is a normally distributed stochastic value with a mean of 1, ranging from 0 to 2; and $l$ is the number of decay steps. $TM_{k,c}$ is the transition matrix that defines whether land use type $k$ is allowed to convert to type $c$ (Verburg & Overmars, 2009). By using this decreasing threshold, the cells with higher overall probabilities are usually most likely to change. CA models with multi-type random patch seeds and threshold descent rules are spatiotemporally dynamic (have temporal consistency), which allows the new land use patches to spontaneously grow and freely develop under the constraints of the growth probabilities.

### 2.3. Generating Sustainable Land Use Structure with MOP

Multiobjective programming (MOP) is an open and flexible method that can incorporate varied ecological and macroeconomic policies (Gardiner & Steuer, 1994). The expectations (e.g., land use diversity) and variables that concern the planners (e.g., food demand) can be taken into account through properly defined objective optimization functions and constraint conditions. This study aims to determine a sustainable land use structure with MOP to support decision making in the study region by employing the objective optimization functions, constraint conditions and parameters proposed by Wang, Li, Zhang, Li, & Zhou (2018). Three optimization objectives were predefined: (1) $max\{f_1(x)\}$, which maximizes the economic benefits; (2) $max\{f_2(x)\}$ which maximizes the ecological service value; and (3) $max\{f_3(x)\}$, which maximizes the ecological capacity. In our study, the optimal sustainable land use structure was assumed to simultaneously maximize these three objectives ($i.e. max\{f_1(x), f_2(x), f_3(x)\}$). Thus, we devised a sustainable development (SD) scenario. In addition, an economic development (ED) scenario that maximized the economic benefits of the land uses ($max\{f_1(x)\}$) and an ecological protection (EP) scenario that maximized the ecological benefits of the land uses ($max\{f_2(x), f_3(x)\}$) were also defined to compare and better illustrate the land use patterns under the SD scenarios. The optimization objectives of MOP are listed in Table 1. The constraint conditions of these objective functions are shown in Table 2.

Table 1. The objective optimization functions of multiobjective programming.

| Function | Formula | Description |
|---|---|---|
| Function for estimating economic benefits. | $f_1(x) = \max \sum_{i=1}^{7} ec_i \cdot x_i$ $= \max\{198.04x_1 + 1.88x_2 + 6.70x_3 + 1831.26x_4 + 0x_5 + 1.69x_6 + 1.88x_7\}$ | The coefficient $ec_i$ is the economic benefits of each land use type (unit: $10^4$ CNY/ha), CNY = Chinese Yuan. |
| Function for estimating ecological service value. | $f_2(x) = \max \sum_{i=1}^{7} esv_i \cdot x_i$ $= \max\{8.56x_1 + 20.63x_2 + 5.80x_3 + 0x_4 + 1.02x_5 + 33.27x_6 + 20.63x_7\}$ | The coefficient $esv_i$ is the ecological service values of each land use type (unit: $10^4$ CNY/ha) |
| Function for estimating ecological capacity. | $f_3(x) = \max \sum_{i=1}^{7} ec_i \cdot x_i$ $= \max\{0.08x_1 + 1.76x_2 + 5.00x_3 + 2.5x_4 + 0x_5 + 9.42x_6 + 1.76x_7\}$ | The coefficient $ec_i$ is the ecological capacity of each land use type. |
| Objective optimization function under the ED scenario | $\max\{f_1(x)\}$ | $x_1 \sim x_7$ represent the area (ha) of grassland ($x_1$), deciduous forest ($x_2$), cropland ($x_3$), urban land ($x_4$), bare land ($x_5$), water area ($x_6$), and evergreen forest ($x_7$). The three optimization objectives share the same constraint conditions. |
| Multiobjective optimization function under the EP scenario | $\max\{f_2(x), f_3(x)\}$ | |
| Multiobjective optimization function under the SD scenario | $\max\{f_1(x), f_2(x), f_3(x)\}$ | |

Table 2. The constraint conditions of the MOP in this study.

| Subject to | Description |
|---|---|
| $\sum_{i=1}^{7} x_i = 479285.19$ (ha) | The sum of the area of all land use types to remain unchanged. |
| $0.55 * (x_1 + x_2 + x_3 + x_7) + 48.93 x_4) \leq 14200000$ | By 2035, the total population is not to be larger than 14.2 million. The population densities of the grassland, cropland, evergreen forest, and deciduous forest are set to 0.55. For urban land, the density value was 48.93 (persons per hectare). |
| $\dfrac{x_5 + x_1}{479285.19} \geq 2.5\%$ | To maintain the land use diversity of the study region, we assumed that the bare land and grassland should account for at least 2.5% of the total area. |
| $\dfrac{0.49 x_1 + x_2 + 0.46 x_3 + x_7}{479285.19} \geq 22\%$ | The coefficients of corresponding land use types denote the 'green equivalent', set according to (Liu, Ming, & Yang, 2002). The total green equivalent was assumed to be larger than 22% by 2035. |
| $x_3 \times 6312 \times 0.4072 \times 2.85 \geq 14200000 \times 209.30 \times 0.187$ | The amount of grain produced by the croplands should not be less than the food demand of the population. In this study, the amount of grain demand per capita was 517.30 kg/capita; the grain self-sufficiency rate was 18.7%; the grain yield per unit cropland area was 6,312 kg/ha; the rate of the crop planting proportion was 40.72%; and the multiple cropping index was 2.85. |
| $0.0066 \leq \dfrac{x_1}{479285.19} \leq 0.0112$ | We set 1.12% (cover proportion of grassland in 2006) as the upper bound and 0.66% (predicted cover proportion of grassland in 2035 by Markov chain) as the lower bound for the percentage of grassland in 2035. |
| $0.2349 \leq \dfrac{x_4}{479285.19} \leq 0.3523$ | The percentage of built-up land will be between 23.49% (80% of the predicted urban land in 2035 by Markov chain) and 35.23% (120% of the predicted urban land in 2035 by Markov chain) in 2035. |
| $0.2502 \leq \dfrac{x_6}{479285.19} \leq 0.373$ | We set 25.02% (predicted cover proportion of grassland in 2035 according to the Markov chain) as the upper bound and 27.30% (cover proportion of grassland in 2013) as the lower bound for the percentage of water area. |
| $0.0136 \leq \dfrac{x_7}{479285.19} \leq 0.0355$ | We set 1.36% (half of the cover proportion of evergreen in 2013) as the upper bound and 3.55% (cover proportion of evergreen in 2006) as the lower bound for the percentage of water area. |
| $0.325 \leq \dfrac{x_7}{x_2} \leq 0.65$ | The proportion between evergreen and deciduous forest was 0.65 in 2013. Under the background of global warming, we assumed that this proportion will not be lower than half of the value in 2013 (0.325) until 2035. |

## 3. Study Area and Data Sources

### 3.1. Study Area

The study area was Wuhan, the capital of Hubei Province, China which covers an area of 8,494.41 km$^2$. The city is located in the intersection of the Yangtze and Han rivers. Water bodies account for a substantial area of Wuhan. Forests are mainly distributed in the hilly areas, and natural vegetation is primarily composed of deciduous broadleaf trees. Wuhan city is a megacity in Central China with a total population of 10.4 million. The gross domestic product (GDP) reached 91.3 billion dollars in 2011, thirteenth among all the cities in China. Wuhan has experienced rapid urban growth in the past decade, which has led to extensive expansion of built-up land (from $4.19 \times 10^4$ ha in 1988 to $49.39 \times 10^4$ ha in 2011) that has encroached upon the surrounding ecologically valuable lands (i.e., cropland, forestland, grassland and water areas). The study area included 11 of the 13 districts in Wuhan, covering the central metropolitan area (Figure 5).

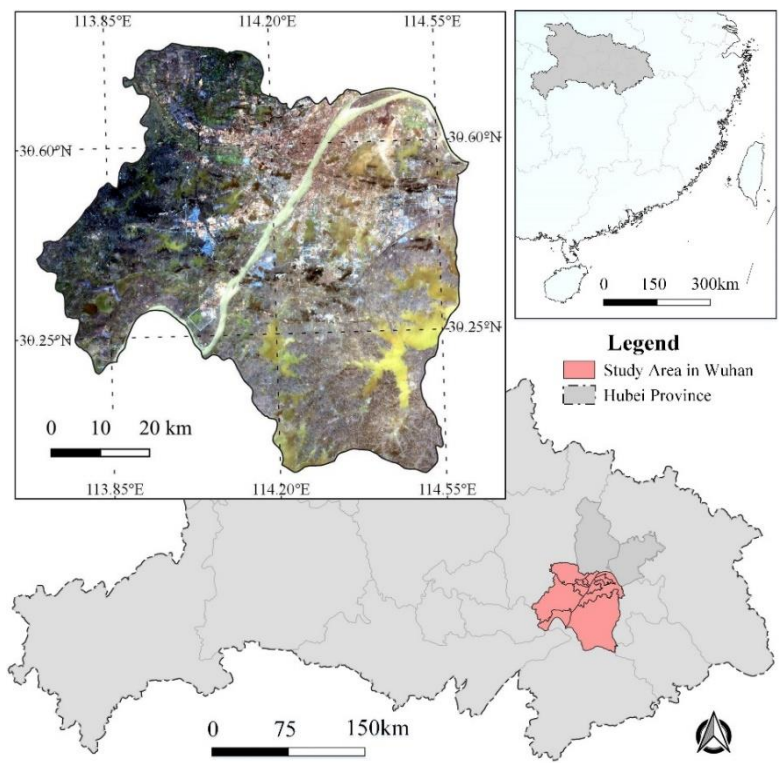

Figure 5. The location and Landsat 8 imagery of the study region in 2013.

### 3.2. Data Sources

The land use data in this study were created by (Liu, Su, Cao, Wang, & Guan, 2019) with an extent of 2931×2931 grid cells and a spatial resolution of 30 m × 30 m and were classified by a decision tree from the corresponding Landsat images in 2003 and 2013. The Landsat imagery was classified into seven land use types, including built-up areas, cropland, broadleaf deciduous forest, broadleaf evergreen forest, grassland, water bodies, and bare land. We also derived raster maps of multiple driving factors (including climatic and environmental drivers and socioeconomic drivers) with the same extent as the land use data. Table 3 lists the data sources of the spatial data used in this study.

Table 3. The spatial driving factors of the land use change in this study.

| Category | Data | Year[1] | Original Resolution | Data resource |
|---|---|---|---|---|
| Land use/cover data | Land use/cover data | 2000-2013 | 30 m | Liu et al., 2019 |
| Socioeconomic driver | Population GDP | 2010 | 1000 m | http://www.geodoi.ac.cn/WebCn/Default.aspx |
| | Proximity to governments | 2013 | 30 m | http://lbsyun.baidu.com |
| | Proximity to highway Proximity to railway Proximity to arterial road Proximity to primary road Proximity to secondary road Proximity to tertiary road | 2015 | 30 m | OpenStreetMap (https://www.openstreetmap.org/) |
| | Proximity to high-speed railway stations | 2013 | 30 m | http://lbsyun.baidu.com/ |
| Climatic and environmental driver | Soil type | 1995 | 1000 m | HWSD v 1.2 (http://westdc.westgis.ac.cn/data/844010ba-d359-4020-bf76-2b58806f9205) |
| | Proximity to open water | | 30 m | CLUD datasets in 2013 |
| | Annual Mean Temperature Annual Precipitation | 1970-2000 | 30 arc-sec | WorldClim v2.0 (http://www.worldclim.org/) |
| | DEM Slope | 2016 | 30 m | NASA SRTM1 v3.0 |

[1] Driving factors collected from different time periods are allowed (Long, Han, Lai, & Mao,2013), but we have made the time periods of the driving factors as close as possible to the time periods of the land use data.

## 4. Model Implementation and Results

The PLUS model was calibrated over the past time interval from 2003 to 2013 to assess the RFC performance and simulation accuracy. Using a random sampling approach, 5% of the sample cells were selected from the spatial variable maps and the land expansion map (Figure 2). These samples served as inputs to the RFC classifier for each increase in a land use type. A total of 14 predictor variables for each tree split and 50 trees were used to build the final RFC models and then to compute the growth probability maps for each land use type. During the calibration time interval, the 3×3 Moore neighborhood was adopted to quantify the neighborhood effects of the PLUS model. The threshold to generate the new land use patches ($\mu_k$) in equation (5) was set to 0.1, and the decay factor of the decreasing threshold $\delta$ in equation (7) was defined as 0.9. The step size ($Step$) of the PLUS model to approximate the land use demand was set to 500. All the parameters are set using a trial-and-error method (Liang et al.,2021).

### 4.1. Model Validation and Comparison

The figure of merit (FOM) was used to validate the simulation results and to compare with those of three other models: the degraded PLUS model (I) (using PAS), the degraded PLUS model (II) (without the multi-type random patch seed function), and the FLUS model (a CA model based on an artificial neural network). The FLUS model has been commonly used to simulate different regions at different scales (regional, continental and global), and can obtain higher simulation accuracies than the other traditional models, such as CLUE-S, ANN-CA and Logistic-CA FLUS (Liu et al.,

2017).

Figure 6 compared the observed land use patterns in 2013 with the simulated land use patterns of the four models. We found that the simulated result of the PLUS model (Figure 6 (d)) was most similar to the observed pattern (Figure 6 (b)) and obtained the highest simulation accuracy (FOM = 0.2642). The c2-3, d2-3, e2-3 and f2-3 panels show incorrect patches simulated by the four simulation models. The incorrect patches derived from the PLUS model are small and uniformly scattered across the study area, which is different from the results generated by the other models that have many obvious and large incorrect patches. The comparison between the simulated accuracy of the PLUS model (Figure 6 (d)) and the degraded PLUS model (I) (Figure 6 (e)) demonstrated that the LEAS had advantages in improving the simulation accuracy (FOM increases from 0.1310 to 0.2642). The simulation accuracy of the PLUS model was higher than that of the degraded PLUS model (II) (Figure 6 (f)), which indicated that the multi-type random patch seed mechanism proposed in this study could help improve the simulation results (FOM increased from 0.2514 to 0.2642). Finally, the comparison between the FOM values of the PLUS model and the FLUS model (0.2642 vs. 0.1895) suggested that the PLUS model was superior to the traditional model in simulating the process of historical land use change.

**Observed 2003** 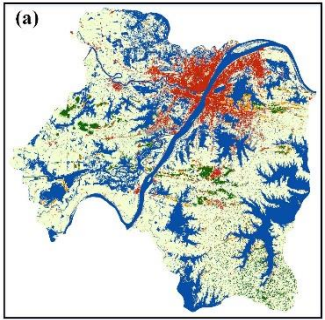 **Observed 2013** 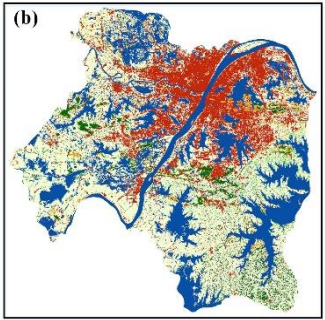

*Legend*
- Grass
- Deciduous forest
- Cropland
- Urban land
- Bare land
- Water area
- Evergreen forest

**FLUS 2013**

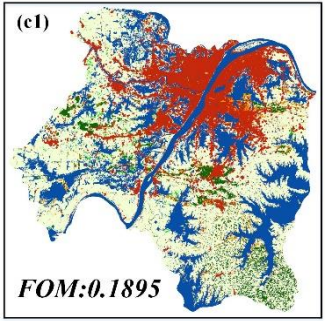 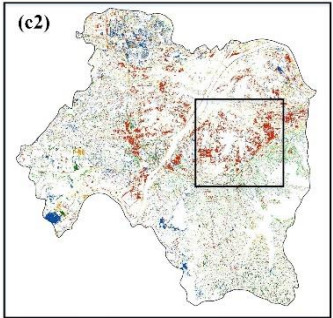 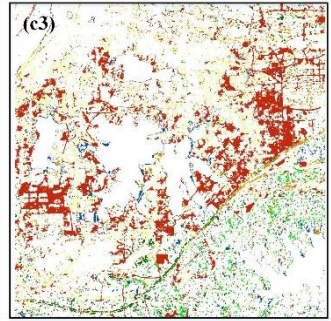

*FOM: 0.1895*

**PLUS 2013**

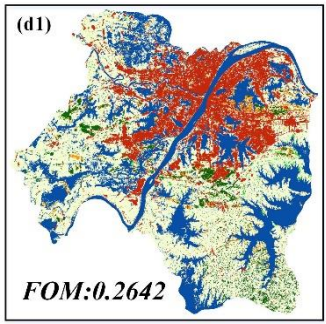 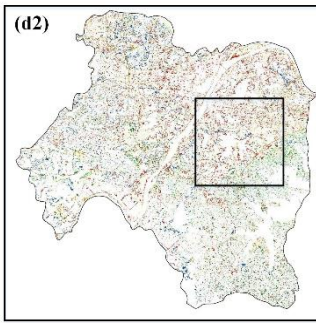 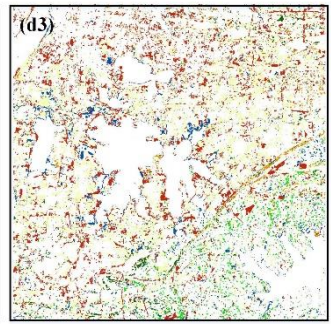

*FOM: 0.2642*

**Degraded PLUS (I) 2013**

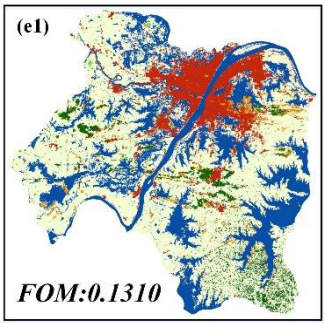 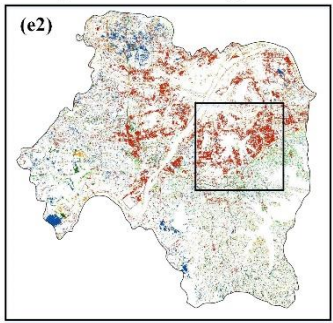 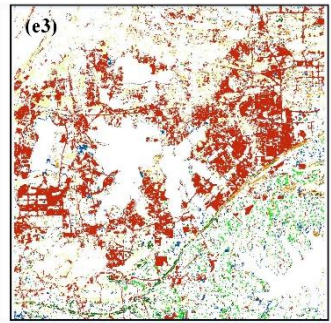

*FOM: 0.1310*

**Degraded PLUS (II) 2013**

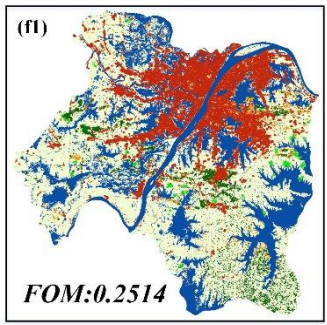 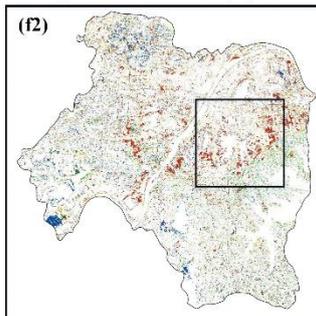 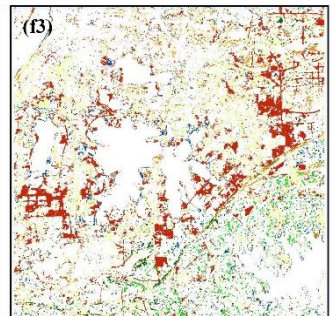

*FOM: 0.2514*

0  15  30  45  60 km          0  5  10  15km

Figure 6. Comparison of the simulated land use patterns using the Degraded PLUS (I) model, Degraded PLUS (II) model, FLUS model and PLUS model. The c2, d2, e2 and f2 panels show the incorrect patches compared to the observed map for 2013 simulated by the four models. The c3, d3, e3 and f3 panels show the incorrect patches of a sub region.

We also validated whether the PLUS model could better trace the change in the landscape patterns compared to other models. The landscape similarity was quantified through a comparison among a series of simulated and observed landscape metrics in 2013. A total of 15 landscape metrics were selected to quantify their landscape similarity: (1) the number of patches (NP); (2) the largest-patch index (LPI); (3-8) the perimeter-area ratio metrics, including mean (PARA_MN), weighted average (PARA_AM), mid-value (PARA_MD), extreme value (PARA_RA), standard deviation (PARA_SD), variable coefficient (PARA_CV); (9-14) Euclidean nearest-neighbor distance metrics include the mean (ENN_MN), weighted average (ENN_AM), mid-value (ENN_MD), extreme value (ENN_RA), standard deviation (ENN_SD), variable coefficient (ENN_CV); and (15) the proportion of like adjacency (PLADJ). The landscape metrics we used are shown in Table 4.

We found that 7 (NP, LPI, PRAR_MD, ENN_AM, ENN_RA, ENN_SD, ENN_CV) among the 15 landscape metrics of the simulated pattern of the PLUS model were closest to the metrics of the observed pattern in 2013, which was higher than the number of closest metrics for the other three models (the FLUS had 5 higher metrics, the degraded PLUS (I) had 3 metrics and the degraded PLUS (II) had 2 metrics). For the other 6 landscape metrics (PRAR_MN, PRAR _AM, PRAR _SD, PRAR _CV, ENN_MD), the simulation by the PLUS model ranked second among the models. These results show that the proposed PLUS model can generate simulation results whose landscape patterns are most similar overall to the observed land use pattern.

Table 4. Landscape similarity of the simulation results of the above models compared with a series of landscape metrics.

| Landscape metrics | NP | LPI | PARA_MN | PARA_AM | PARA_MD | PARA_RA | PARA_SD | PARA_CV | ENN_MN | ENN_AM | ENN_MD | ENN_RA | ENN_SD | ENN_CV | PLADJ |
|---|---|---|---|---|---|---|---|---|---|---|---|---|---|---|---|
| **Degraded PLUS (II)** | *65606[2]* | 22.288 | 1046.828 | <u>**149.293[1]**</u> | *1111.111[2]* | 1291.287 | 326.421 | 31.182 | *149.607[2]* | 74.428 | *84.853[2]* | 9991.075 | *205.391[2]* | 137.287 | <u>**88.803[1]**</u> |
| **Degraded PLUS (I)** | 52320 | 27.104 | 990.048 | 115.302 | <u>**1000[1]**</u> | <u>**1296.911[1]**</u> | 340.334 | 34.375 | <u>**150.954[1]**</u> | *65.497[2]* | *84.853[2]* | 8125.658 | 239.282 | 158.513 | 91.352 |
| **FLUS** | 46804 | *21.019[2]* | <u>**940.227[1]**</u> | 126.59 | <u>**1000[1]**</u> | *1295.291[2]* | <u>**364.202[1]**</u> | <u>**38.736[1]**</u> | 181.292 | 74.921 | <u>**94.868[1]**</u> | *7867.78[2]* | 239.937 | *132.349[2]* | 90.506 |
| **PLUS** | <u>**71063[1]**</u> | <u>**20.418[1]**</u> | *964.436[2]* | *144.957[2]* | <u>**1000[1]**</u> | 1294.482 | *348.534[2]* | *36.139[2]* | 146.487 | <u>**66.917[1]**</u> | *84.853[2]* | <u>**4713.301[1]**</u> | <u>**179.391[1]**</u> | <u>**122.462[1]**</u> | *89.128[2]* |
| **Observed 2013** | 77890 | 20.064 | 897.573 | 152.795 | 888.889 | 1300.995 | 359.05 | 40.002 | 163.383 | 69.791 | 108.167 | 5389.853 | 174.62 | 106.878 | 88.54 |

[1] The underline and bold values are the first closest landscape metrics to the observed land use pattern.

[2] The italic and bold values are the second closest landscape metrics to the observed land use pattern.

**4.2. Analyzing the Underlying Driving Forces of the LULC with LEAS**

By adopting the LEAS proposed in this study, the analysis of the driving factors for land use change can be more rigorous and have a more clear meaning than those in previous studies (Gounaridis, Chorianopoulos, Symeonakis, & Koukoulas, 2019;Zhang et al., 2019). For example, a previous CA model based on a transition analysis strategy trained 18 random forest models and derived corresponding 'from-to' transition probabilities; but other possible transitions were ignored (Gounaridis, Chorianopoulos, Symeonakis, & Koukoulas, 2019). Although this process can drive the allocation of cells in CA, the contribution of multiple driving factors to the individual land use types cannot be derived from the training step. The 'from' land use type was specified in the training process of each random forest model. Different than the TAS, the LEAS used by the PLUS model merges all the 'from' land use types of a "to" land use type. Thus, the significance of the measurement of variable importance was more explicit, which could be understood as the contributions of multiple driving factors in driving the other land uses to convert to a specified target land use type. Figure 6 shows the variable importance that informs the contribution of each variable to the growth of the three land use types.

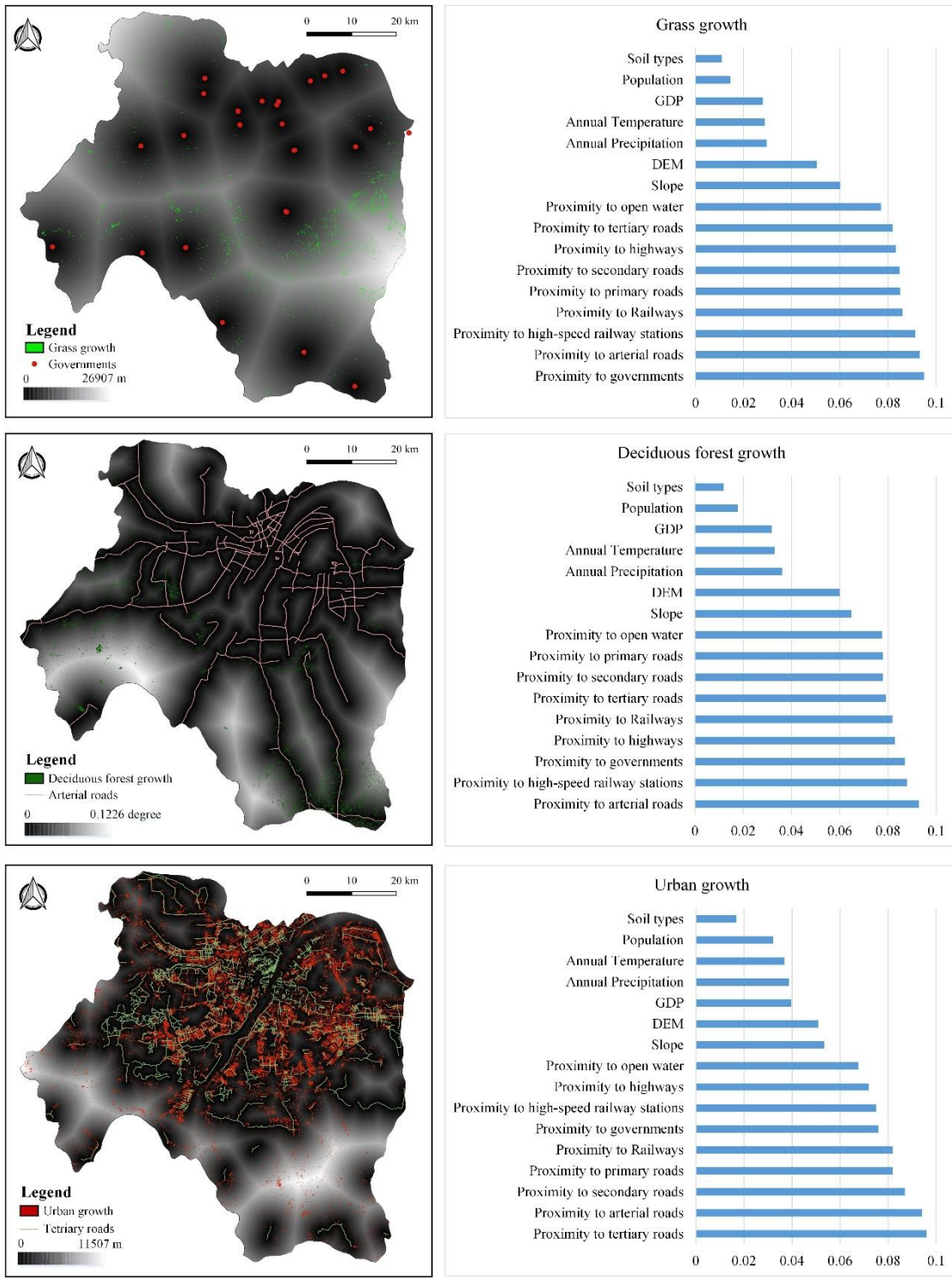

Figure 7. The importance of the contribution of each variable to the growth of three land use types. The most important factors were overlapped with the expansion of the corresponding land uses.

For the grasslands, we found that proximity to administrative centers (governments) had the most influence on the growth of grass. We overlaid the grass growth with the proximity to administrative centers and found that the new grassland was mainly distributed in areas far away from the governments. The areas around the governments were generally areas where human beings are most active. This result indicates that grass was most likely to grow in areas where it was not strongly impacted by human activities. In contrast to the grassland, deciduous forest was most influenced by proximity to the arterial roads. Although almost no new deciduous forests grew in the metropolitan areas where the density of arterial roads was relatively high from 2003 to 2013, the new deciduous forests were most likely to grow in the regions adjacent to arterial roads in the suburbs. This suggested that the growth of deciduous forest may have been managed by human beings. Policymakers tend to plant new deciduous forests along the arterial roads that go through the suburbs. Moreover, we also found that urban growth was most influenced by tertiary roads. The distribution of the new urban growth also had a high consistency with the pattern of the tertiary roads. This is hardly surprising, as most new urban growth implies and depends on extending first the local, then the connector road networks.

### 4.3. Forecasting the Different LULCs under Multiple Scenarios

For future simulations, we created one baseline scenario and three alternative future scenarios using a Markov chain model and multiobjective programming (MOP), respectively. Then, the PLUS model was employed to allocate the predicted land use demand at the local scale of land use change at a fine resolution to support the master plan of Wuhan (2017–2035). The main differences between the four scenarios were the future demands. Other parameters of the four scenarios were set the same as in the validation process discussed in section 4.

### 4.3.1. Scenario Description

Based on the optimization objectives and constraint conditions, we solved the multiobjective optimization problems using Lingo 12.0 software. Table 5 shows the predicted land use demands in 2035 under the different scenarios. In addition, a baseline scenario (BS) was derived from a Markov chain (Bai et al.,2018), which represents the historical trend of land use change in Wuhan. A transfer probability matrix for 2003–2013 was generated with the Markov chain model to predict land use demands in 2026 and 2039. Next, linear interpolation was applied to project the land use demands in 2035. We found that under the ED and SD scenarios, the decrease in cropland was the same, but the areas of water, evergreen forest, and deciduous forest were greater in the SD scenario. This comparison indicated that urban expansion avoids encroaching upon forest land and water areas and can promote more sustainable regional development. The EP scenario has the most forestland, grassland, water area, and cropland, and the least amount of urban land among these scenarios. It is worth noting that the BS scenario will lose the most cropland and grassland, which indicates that more attention should be paid to protecting the cropland and grassland under the current trend of land use change.

Table 5. Land use demands under different scenarios (unit: ha).

| Type | | Markov | Multiobjective Optimization | | |
|---|---|---|---|---|---|
| | | BS scenario | ED scenario | EP scenario | SD scenario |
| | 2013 | 2035 | 2035 | 2035 | 2035 |
| Grassland | 3995.64 | 3178.99 | 5367.99 | 5367.99 | 5367.99 |
| Deciduous forest | 20801.16 | 25010.82 | 10033.91 | 27499.03 | 22814.06 |
| Cropland | 222968.79 | 175295.11 | 187522.29 | 188970.75 | 187522.29 |
| Urban land | 83651.49 | 140757.93 | 149542.20 | 112584.06 | 130766.13 |
| Bare land | 3547.35 | 3020.94 | 383.43 | 383.43 | 383.43 |
| Water area | 130827.87 | 119935.84 | 119917.08 | 130844.88 | 124988.76 |
| Evergreen forest | 13570.38 | 12085.56 | 6518.28 | 13635.01 | 7442.53 |

4.3.2. Projecting Future LULCs with the PLUS Model

Driven by future land use demands, the PLUS model was applied to simulate land use changes under the four different scenarios. The difference maps of the simulation results in 2035 are shown in Figure 8. The main characteristics of land use change under the BS and ED scenarios were the rapid urban expansion and the reduction of cultivated land compared to the actual land use in 2013, but the new urban patches under the ED scenario are more compact. The land expansion map under the ED scenario is characterized by the emergence of many large deciduous forest and grassland patches. The urban growth and expansion of deciduous forest under the SD scenario are intermediate between the patterns under the ED and EP scenarios.

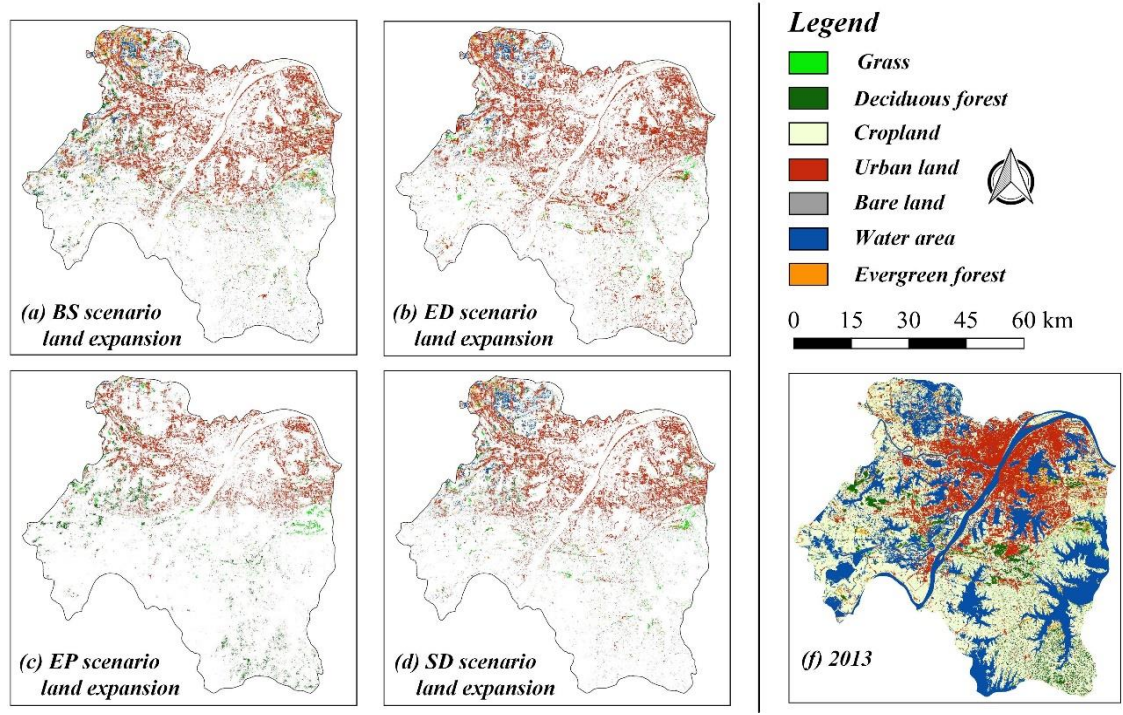

Figure 8. The land expansion maps from 2013-2035 generated by the PLUS model under different scenarios.

In a sub-region around Tangxun Lake in Wuhan (Figure 9), the urban growth under the BS scenario will expand and encroach on a large portion of the cropland, but the largest deciduous forest patch would be well preserved. The deciduous forest corridor would become larger and more continuous, which was different from the situation in the ED scenario. Although the urban growth in the ED scenario was much more compact, it preserved more cropland, which could create more economic benefits at the cost of encroaching on more deciduous forest. Thus, the deciduous forest patches would be much smaller, and the deciduous forest corridor would almost disappear under the ED scenario. In contrast to the ED scenario, the EP scenario had the highest demand for forestland and the best-preserved deciduous forest patches among all the scenarios. The deciduous forest corridor under the EP scenario had the most obvious growth. Under the SD scenario, the urban growth would be controlled to reduce the influence on the deciduous forest. The distribution pattern of the deciduous forest remained almost the same as in 2013. A new evergreen patch was predicted to appear in the center of the largest deciduous forest patch.

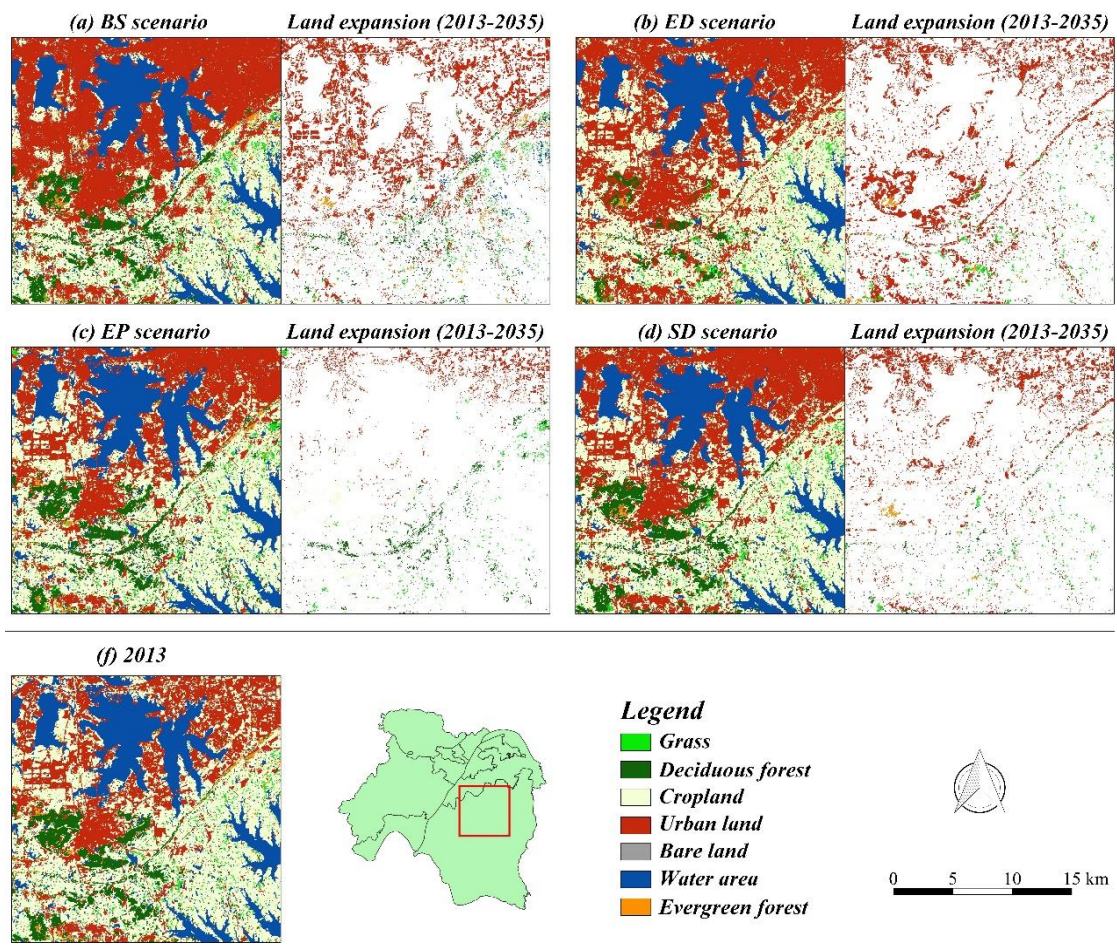

Figure 9. A local-scale subset of the simulated land use patterns in 2035 and land expansion maps from 2013 to 2035 under different scenarios.

## 5. Discussion

The CA module with the multi-type patch strategy and the rule-mining framework based on the land expansion analysis strategy were closely linked and together constitute the PLUS model. The training process of the PLUS model can directly provide quantitative information on how the various driving factors influenced the expansion of the multiple land use types (Figure 7). Considering that the forces of land use change may change with time (i.e., variation in the driving factors), the transition rules obtained from the training process of the PLUS were more valuable and flexible than the distribution rules that have been mined in previous studies, because the transition rules in this study have a temporal aspect, which gives transition rules the ability to describe the properties of land use change for specific time interval (Figure 3)(He et al., 2020). This advantage can help policymakers understand how the driving factors (e.g., growth in arterial roads) influence short-term land use change. In addition, as indicated by the results of the model validation and comparison, the simulation results obtained higher simulation accuracy (FOM increased from 0.1895 to 0.2642) and gave more similar landscape metrics than the other models tested (Table 4). Thus, more reliable simulation results under the different future scenarios can be expected with the PLUS model.

By using the MOP, policymakers can obtain optimized land use patterns under different policies and scenarios. For example, the optimal land use structures for economic development were calculated under the ED scenario. The total economic benefits under this scenario were expected to reach $2.764 \times 10^{12}$ CNY. Similarly, a land use structure that could balance both the ecological service values and the ecological capacity was also obtained under the EP scenario, the values of which were $6.3657 \times 10^{10}$ CNY and $2.6115 \times 10^{10}$, but the economic benefits were predicted to decrease to $2.088 \times 10^{12}$ CNY. These two scenarios were adopted by Wang et al.,(2018), and we projected the land use structures in our study region to 2035 under these two scenarios.

Moreover, previous studies have not arranged scenarios to find the balance between the economic benefits and ecological benefits. This study designed an SD scenario aimed at maximizing the three aforementioned objectives simultaneously. The three benefit values found by the MOP were $2.4205 \times 10^{12}$ CNY (economic benefits), $5.9165 \times 10^{10}$ CNY (ecological service value), and $2.4956 \times 10^{10}$ (ecological capacity). The land use structure that helped to improve regional sustainable development was obtained (Table 5), which is also of importance for assisting policymakers in determining future management objectives and reasonable land polices (Cao, Huang, Wang, & Lin, 2012).

## 6. Conclusion

This research has presented a rule-mining framework based on a land expansion analysis strategy (LEAS), which can bring new scientific understanding of LULC to a study region. The LEAS simplified the analysis of land use change while maintaining the ability to support multi-type, complex land use change. Based on the growth probabilities output by the LEAS, a CA model based on multi-type random patch seeds (CARS) was proposed, to better simulate the patch growth of multiple land use types at a fine-scale resolution. As a result, a more realistic landscape pattern can be generated to support decision-making.

By applying a combination of the LEAS and CARS, we constructed a patch-generating land use simulation (PLUS) model that was available to simulate the change of land use patches and to analyze the underlying drivers of land use dynamics. The PLUS model was calibrated using a simulation of Wuhan (2003-2013) and obtained higher simulation accuracy and more similar landscape patterns than the other models tested. The variable importance derived from the LEAS revealed some transition effects that could not be found by the previous analysis methods. For example, grassland was most likely to grow in areas where it was not strongly impacted by human activities, and new deciduous forests were most likely to grow along arterial roads in the suburbs.

By coupling the PLUS model with the MOP, the land use structures under the different scenarios in 2035 and their corresponding future economic and ecological benefits can be predicted, which are of great importance for policymakers to plan for the future development goals of the study region. In particular, the land use structure under the SD scenario can be regarded as a baseline for examining whether the study region has developed along a sustainable pathway. In summary, the PLUS model produced results that were more accurate, had more reliable landscape patterns and allowed for important insights concerning the drivers of land expansion. It also can produce substantive guidance for policymakers on how to manage future land use patterns with different development objectives. We encourage others to use this approach for understanding the mechanisms of land expansion and to obtain optimal land use patterns under different policies.